\documentclass[twocolumn,showpacs,unsortedaddress,superscriptaddress,showkeys,nofootinbib,preprintnumbers,letterpaper]{revtex4-1}
\usepackage{acronym}
\usepackage{amsmath}
\usepackage{amssymb}
\usepackage{graphics}
\usepackage{subfigure}
\usepackage[squaren]{SIunits}
\usepackage[usenames]{color}

\newcommand*{\diff}{\,\mathrm{d}}
\newcommand{\abs}[1]{\left\lvert #1 \right\rvert}
\newcommand{\chisq}{\chi^{2}}
\newcommand{\Deff}{D_{\mathrm{eff}}}
\newcommand{\Dh}{{D_{\mathrm{H}}}}
\newcommand{\Dnl}{\tilde{D}}
\newcommand{\figref}[1]{FIG.\ \ref{#1}}
\newcommand{\lr}{\mathcal{L}}

\newcommand{\Msun}{M_{\odot}}
\newcommand{\noise}{\mathrm{noise}}

\newcommand{\signal}{\mathrm{signal}}
\newcommand{\sigparams}{\bar{\theta}}
\newcommand{\snr}{\rho}

\addunit{\annum}{a}
\addunit{\parsec}{pc}

\DeclareMathOperator{\order}{O}

\begin{document}


\title{Likelihood-Ratio Ranking Statistic for Compact Binary Coalescence
Candidates with Rate Estimation}

\author{Kipp Cannon}
\email{kipp.cannon@ligo.org}
\affiliation{Canadian Institute for Theoretical Astrophysics, 60 St.\ George Street, University of Toronto, Toronto, ON, M5S 3H8, Canada}

\author{Chad Hanna}
\email{chad.hanna@ligo.org}
\affiliation{Dept.\ of Physics, Pennsylvania State University, 104 Davey Lab \#253, University Park, PA 16802}

\author{Jacob Peoples}
\email{jacob.peoples@queensu.ca}
\affiliation{School of Computing, 557 Goodwin Hall, Queen's University, Kingston, ON, K7L 2N8}

\begin{abstract}
We present a new likelihood-ratio ranking statistic for use in searches for
gravitational waves from the inspiral and merger of compact object
binaries.  Expanding on previous work, the ranking statistic incorporates a
model for the correlations in the signal-to-noise ratios with which signals
will be seen in a network of ground-based antennas while retaining an
algebraic procedure for mapping ranking statistic values to false-alarm
probability.  Additionally, the ranking statistic enables the
implementation of a rigorous signal rate estimation technique.  We
implement the ranking statistic and demonstrate its use including signal
rate estimation in an analysis of a simulated signal population in
simulated noise.
\end{abstract}

%
%
%
%


\maketitle
\acrodef{BNS}{binary neutron star}
\acrodef{CBC}{compact binary coalescence}
\acrodef{CCDF}{complementary cummulative distribution function}
\acrodef{CDF}{cummulative distribution function}
\acrodef{CW}{continuous wave}
\acrodef{DOF}{degree-of-freedom}
\acrodef{FAR}{false alarm rate}
\acrodef{FAP}{false alarm probability}
\acrodef{GRB}{gamma-ray burst}
\acrodef{GW}{gravitational-wave}
\acrodef{LR}{likelihood ratio}
\acrodef{LVC}{LIGO Scientific and Virgo Collaborations}
\acrodef{PDF}{probability density function}
\acrodef{RV}{random variable}
\acrodef{SNR}{signal-to-noise ratio}
\acrodef{TAP}{true-alarm probability}

\section{Introduction}
\label{section3}

An essential ingredient of searches for gravitational waves is a statistic
by which to rank candidates from least signal-like to most signal-like.
The \ac{LR} has been shown to provide the most powerful detection statistic
at fixed false-alarm probability \cite{neymanpearson:1933}, but due to
technical challenges all ranking statistics used to date in searches for
\acp{CBC} have been approximations of a \ac{LR}.  To be useful, the ranking
statistic must not only be effective at ranking signals above noise, but it
must not be computationally costly to implement, and there must be a known
mapping from ranking statistic value to false-alarm probability.

In \cite{ruslan2012a}, Biswas \textit{et al.}\ developed theoretical
aspects of the use of the \ac{LR} \cite[Section 7.6]{winklerhays} as the
ranking statistic for searches for \acp{GW} from \acp{CBC}.  For example,
they showed that when a \ac{LR}-based ranking statistic is used the
detection efficiency is not diminished when the volume of the parameter
space over which a search is conducted is increased, in contrast to
techniques using other ranking statistics \cite{broek2009}.  That work was
preceded by an earlier effort to incorporate the Virgo antenna
\cite{Virgo2012} into a search for \acp{CBC} \cite{s5vsr1ruslan}, wherein
the substantial difference in sensitivity of the Virgo antenna next to the
three LIGO antennas \cite{LIGO2009} of the day motivated the need to
explicitly account for the relative sensitivities of all antennas in the
ranking statistic.  Biswas \textit{et al.}\ demonstrated an implementation
of a \ac{LR} ranking statistic on LIGO's S4 data, but it was implemented as
a rescaling of the traditional ``combined effective \ac{SNR}'' ranking
statistic \cite[equation (26)]{ruslan2012a}, and so while the \ac{LR}
allowed events from times when different instruments were operating or from
different regions of mass parameter space to be ranked on a common scale it
otherwise provided no information not already contained in the combined
effective \ac{SNR}, and they continued to rely on time-shift analyses
\cite{babak2012a} to map \ac{LR} values to \ac{FAP}.  In the related work
\cite{ruslan2012b}, Biswas \textit{et al.}\ demonstrate the use of the
\ac{LR} to combine the results from several, possibly degenerate or
uninformative, searches for \acp{GW} into a single unified search result.

In \cite{s56vsr123strings}, Aasi \textit{et al.}\ built upon the
histogram-based \ac{LR} ranking statistic described in \cite{Cannon2008}
for coincidences in a time-frequency burst search to develop a \ac{LR}-like
ranking statistic for searches for \ac{GW} bursts from cosmic string cusps.
Subsequently, in \cite{cannon2012b} Cannon \textit{et al.}\ expanded upon
that technique to develop an histogram-based \ac{LR}-like ranking statistic
for searches for \acp{GW} from \acp{CBC}, and demonstrated its use.  There
has also been work to approximate \ac{LR} ranking statistics using
space-partitioning decision trees in searches for \ac{GW} bursts
\cite{PhysRevD.88.062006}, and in \ac{CBC} searches
\cite{bakercaudillhodge:2014};  in the latter the mapping from ranking
statistic to \ac{FAP} was accomplished through the use of time-slides while
in the former the ranking statistic was applied to \ac{GRB}-triggered
searches and the mapping was constructed using on-source/off-source
comparisons.  In contrast, the ranking statistic developed in
\cite{cannon2012b} had the property that if applied to the candidates
collected from a search pipeline possessing certain simplifying
characteristics, it was possible to compute the mapping from ranking
statistic to \ac{FAP} algebraically.

The ranking statistic in \cite{cannon2012b} was an approximation of the
\ac{LR} in which it was assumed that the characteristics of a signal-like
event in one instrument were independent of that same event's
characteristics in other instruments --- that the \ac{PDF} in the ranking
statistic's numerator could be factored into a product of single-instrument
terms.  This approximation simplified the software needed to implement the
ranking statistic and afforded a simple numerical integration scheme for
computing the \ac{FAP} for any value thereof.

Although the ability to compute \acp{FAP} algebraically was a significant
improvement over previous techniques, three limitations remained.  (i) The
ranking statistic assumed the \acp{SNR} at which a signal is observed in
the various instruments are independent random variables.  For low
\acp{SNR} where plausible early detections are expected this is a good
approximation but for higher \acp{SNR} this approximation yields a
less-than-optimal ranking statistic in that it does not properly penalize
candidates with implausible \ac{SNR} combinations that have resulted from,
\textit{e.g.}, non-stationary noise in one detector.  (ii) No attempt was
made to include within the ranking statistic knowledge of which instrument
combination had identified the candidate, leading to a unique ranking
statistic scale for each combination of instruments, something the work of
Biswas \textit{et al.}\ had already addressed.  (iii) Finally, although the
\ac{FAP} for each value of the ranking statistic could be computed, with
only an approximate \ac{PDF} for the numerator a \ac{TAP} (probability of
at least one candidate at or above a given value of the ranking statistic
given the presence of a signal) could not be computed which excluded the
possibility of combining the ranking statistic with the rate estimation
technique of Farr \textit{et al.}\ \cite{farr2013a}.

Here, we address these limitations by showing how to compute and
incorporate the probability of observing signals and noise events in
various combinations of instruments into both the numerator and the
denominator, and how to incorporate the joint \ac{PDF} for the \acp{SNR} in
the numerator.  These modifications will require us to abandon the use of
explicit numerical integration for computing the ranking statistic \ac{PDF}
in signal-free data --- the first step in mapping ranking statistic values
to \acp{FAP} --- and switch to an importance-weighted sampling procedure,
the details of which will also be explained.

\section{Example}

\begin{figure}
\resizebox{\linewidth}{!}{\includegraphics{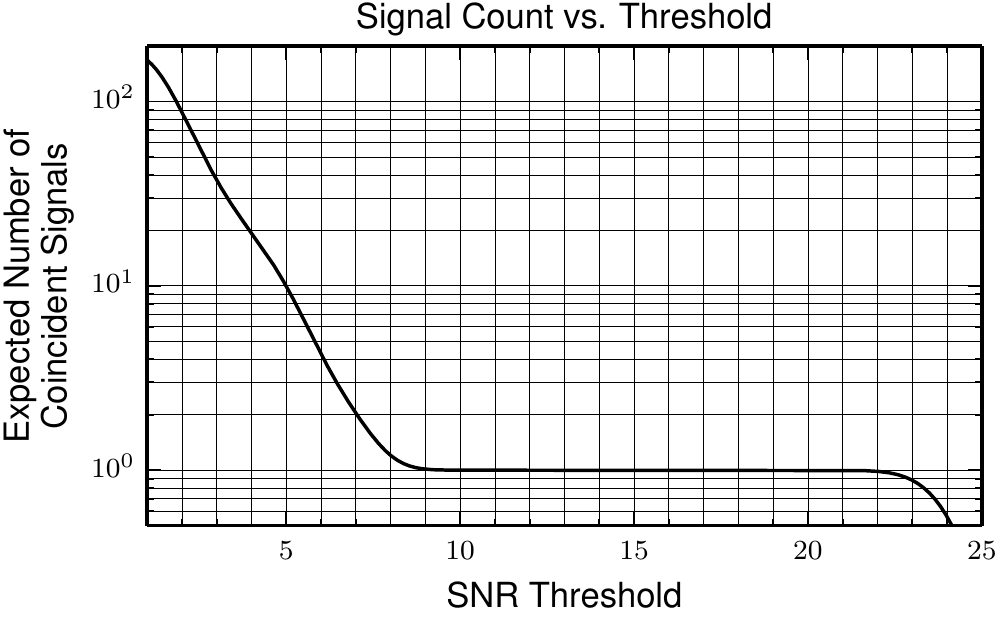}}
\caption{Expectation value for the number of coincident candidates to be
recovered from the signal population injected into test data set as a
function of the single-detector \ac{SNR} threshold applied in the search
assuming an \ac{SNR} recovery efficiency of 0.975.}
\label{fig7}
\end{figure}

To demonstrate the application of the ranking statistic, we have performed
a test analysis of about \(\unit{5.23 \times 10^{6}}{\second}\) of
stationary Gaussian noise simulating the Advanced LIGO and Virgo antennas
and possessing the spectral densities described as the ``early'' noise
curves in \cite{LIGO-T1200307,LIGO-T1300121}.  A population of \ac{BNS}
merger events was injected into the data from a population of sources
distributed uniformly in volume to a distance of
\(\unit{300}{\mega\parsec}\), uniformly in component masses in the range
\(\unit{1}{\Msun}\)--\(\unit{3}{\Msun}\), with Gaussian-distributed
dimensionless spins centred on 0 with a standard deviation of 0.4 a cut-off
of 0.7 and isotropic orientations, and occurring at a mean rate of
\(\unit{10}{\mega\parsec^{-3}\mega\annum^{-1}}\).  This is not meant to be
astrophysically realistic \cite{0264-9381-27-17-173001}, but to provide a
useful sample of detectable signals.  Comparing the injected waveforms to
the noise spectra and accounting for the relative orientations of the
antennas and the sources at the time of each simulated event, one can
determine the \ac{SNR} at which the signal should be seen in each antenna,
and from that estimate how many signals should be seen in two or more
antennas (\textit{i.e.}, form a coincident candidate event) as a function
of the single-detector \ac{SNR} threshold applied.  This is shown in
\figref{fig7}.
The data was filtered using a template bank consisting of 6278 non-spinning
\ac{BNS} waveforms with component masses in the range
\(\unit{1}{\Msun}\)--\(\unit{3}{\Msun}\), and having a minimal match of
0.975 (with respect to the template family).  The template bank was laid
out using the \texttt{lalapps\_tmpltbank} program from the LALSuite
software package \cite{LALSuite}, and candidate events collected using the
\texttt{gstlal\_inspiral} program from the GstLAL software package
\cite{GstLAL,cannon2011b}.  The plots and results shown in what follows are
taken from this analysis.

\section{Ranking Statistic, \(\lr\)}
\label{section4}

The ranking statistic is the \ac{LR}
\begin{widetext}
\begin{multline}
\label{eqn1}
\lr\left(\left\{\Dh_{\mathrm{H1}}, \Dh_{\mathrm{L1}}, \ldots\right\},
\left\{\mathrm{H1}, \mathrm{L1}, \ldots\right\}, \snr_{\mathrm{H1}},
\chisq_{\mathrm{H1}}, \snr_{\mathrm{L1}}, \chisq_{\mathrm{L1}}, \ldots,
\sigparams\right) \\
   = \lr\left(\ldots | \sigparams\right) \lr\left(\sigparams\right)
   = \frac{P\left(\left\{\Dh_{\mathrm{H1}},
   \Dh_{\mathrm{L1}}, \ldots\right\}, \left\{\mathrm{H1}, \mathrm{L1},
   \ldots\right\}, \snr_{\mathrm{H1}}, \chisq_{\mathrm{H1}},
   \snr_{\mathrm{L1}}, \chisq_{\mathrm{L1}}, \ldots | \sigparams,
   \signal\right)}{P\left(\left\{\Dh_{\mathrm{H1}}, \Dh_{\mathrm{L1}},
   \ldots\right\}, \left\{\mathrm{H1}, \mathrm{L1}, \ldots\right\},
   \snr_{\mathrm{H1}}, \chisq_{\mathrm{H1}}, \snr_{\mathrm{L1}},
   \chisq_{\mathrm{L1}}, \ldots | \sigparams, \noise\right)}
   \lr\left(\sigparams\right)
\end{multline}
\end{widetext}
where \(\left\{\Dh_{\mathrm{H1}}, \Dh_{\mathrm{L1}}, \ldots\right\}\) is
the set of horizon distances for all instruments in the network at the time
the event is observed (see Appendix \ref{appendix2}), \(\left\{\mathrm{H1},
\mathrm{L1}, \ldots\right\}\) is the specific set of instruments in which
the event has been observed, \(\snr\) and \(\chisq\) are the template
matched-filter \acp{SNR} and \(\chisq\) values \cite[and references
therein]{allenchi2} for the candidate in each of those instruments, and
\(\sigparams\) are the intrinsic parameters of the template.  The time at
which the event is observed enters implicitly through the horizon distances
which fluctuate in time as environmental factors and operating conditions
at the observatories change; in fact, \(\left\{\Dh_{\mathrm{H1}},
\Dh_{\mathrm{L1}}, \ldots\right\}\) is best thought of as simply a clock,
and since time influences the distributions of other parameters through
their dependence on the (time-dependent) instrument sensitivities, the
labelling of times on the clock is most conveniently done with the
instrument sensitivities themselves.  We'll say a few more words about the
time dependence and the ranking statistic below.  The template's intrinsic
parameters, \(\sigparams\), can be a label identifying a template in the
search, or identifying a group of templates if that is found to be
convenient.  Anticipating what follows, we have factored \(\sigparams\) out
of the \acp{PDF} on the right-hand side of \eqref{eqn1}.  For brevity we
will drop \(\sigparams\) from the notation in the remainder of Section
\ref{section4};  it should be remembered that we will be deriving \acp{PDF}
\emph{for a choice of \(\sigparams\)}.

The differences between the times at which the event is observed at each
location on Earth are not used, nor are the phases with which the waveforms
are recovered at each observatory.  These omissions are an obvious avenue
for future improvements but, for signals, the inter-antenna time delays and
waveform phases are correlated with the \acp{SNR} and their inclusion would
make the numerator substantially more difficult to compute numerically.

\begin{figure}
\resizebox{\linewidth}{!}{\includegraphics{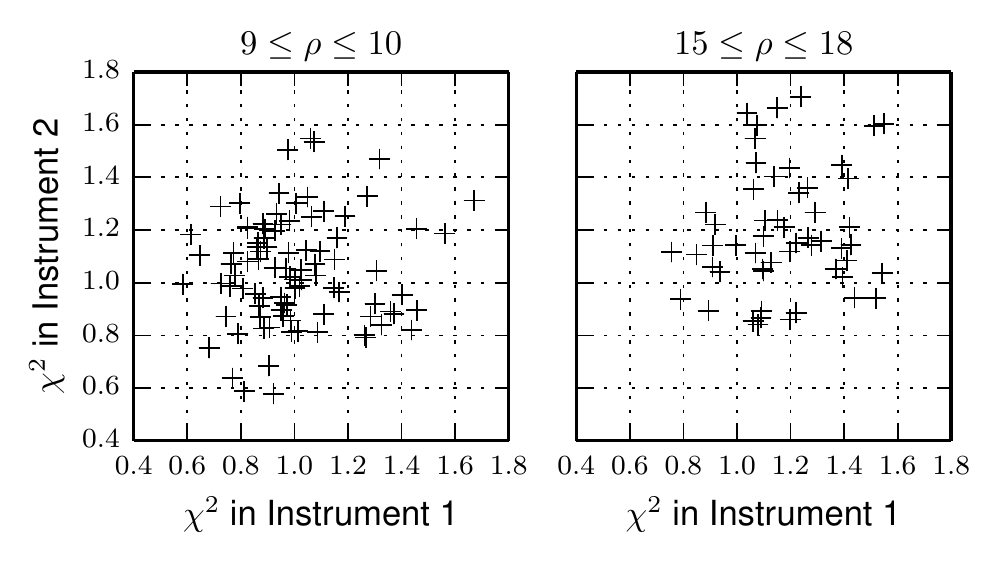}}
\caption{Statistical independence of \(\chisq\) observed in different
instruments.  Plots show reduced \(\chisq\) values from distinct (but
irrelevant) instruments in coincidences arising from software injections
subject to the constraint that the observed \acp{SNR} be in the given
ranges.  There is a trend towards larger \(\chisq\) values as the \ac{SNR}
of the injection is increased, but at a given \ac{SNR} the \(\chisq\)
values in different instruments are not significantly correlated indicating
that instrument noise dominates the residuals over template/signal
mismatch.  Note that this is despite the search being conducted with
non-spinning templates while the injections have spin and include
orbital precession.}
\label{fig8}
\end{figure}

We factor the numerator of the fraction in \eqref{eqn1} as
\begin{widetext}
\begin{multline}
\label{eqn2}
   = P\left(\left\{\Dh_{\mathrm{H1}}, \Dh_{\mathrm{L1}},
   \ldots\right\}\right) P\left(\left\{\mathrm{H1}, \mathrm{L1},
   \ldots\right\} | \left\{\Dh_{\mathrm{H1}}, \Dh_{\mathrm{L1}},
   \ldots\right\}, \signal\right) \\\times P\left(\snr_{\mathrm{H1}},
   \snr_{\mathrm{L1}}, \ldots | \left\{\Dh_{\mathrm{H1}},
   \Dh_{\mathrm{L1}}, \ldots\right\}, \left\{\mathrm{H1}, \mathrm{L1},
   \ldots\right\}, \signal\right) \prod_{\mathrm{inst} \in
   \left\{\mathrm{H1}, \mathrm{L1}, \ldots\right\}}
   P\left(\chisq_{\mathrm{inst}} | \snr_{\mathrm{inst}}, \signal\right).
\end{multline}
\end{widetext}
That is, we write it as the probability of the collection of instruments
having some given sensitivities multiplied by the probability that a
(detectable) signal is visible in a set of those instruments (and no
others) given the sensitivities of all instruments multiplied by the joint
probability density of the \acp{SNR} in that set of instruments multiplied
by the product of the probability densities for the \(\chisq\) values given
the \acp{SNR} and the instruments.  In particular, we assume here that
except for their correlation with \ac{SNR} the \(\chisq\) values observed
in different instruments for the same event are statistically independent
random variables, \textit{i.e.}, that the residual from which the
\(\chisq\) is computed is dominated by instrumental noise and not by
systematic waveform errors (which would be common to all instruments in the
network).  The validity of this assumption can be checked visually with
scatter plots like those shown in \figref{fig8}.

We assume the noise processes in the different instruments are
statistically independent of each other and independent of the sensitivity
of the antennas to \acp{GW} and so the denominator of the fraction in
\eqref{eqn1} can be factored into several low-dimensional terms.
\begin{multline}
\label{eqn3}
   = P\left(\left\{\Dh_{\mathrm{H1}}, \Dh_{\mathrm{L1}},
   \ldots\right\}\right) P\left(\left\{\mathrm{H1}, \mathrm{L1},
   \ldots\right\} | \noise\right) \\\times \prod_{\mathrm{inst} \in
   \left\{\mathrm{H1}, \mathrm{L1}, \ldots\right\}}
   P\left(\snr_{\mathrm{inst}}, \chisq_{\mathrm{inst}} | \noise\right).
\end{multline}

The factor \(P\left(\left\{\Dh_{\mathrm{H1}}, \Dh_{\mathrm{L1}},
\ldots\right\}\right)\) is common to the numerator and denominator and, so,
for the purpose of evaluating \(\lr\) the horizon distances appear only
parametrically and only in the numerator.  Later, when using the numerator
and denominator to map \(\lr\) values to \ac{FAP} and \ac{TAP} the factor
\(P\left(\left\{\Dh_{\mathrm{H1}}, \Dh_{\mathrm{L1}},
\ldots\right\}\right)\) will be included in those integrals.  Since we
understand \(\left\{\Dh_{\mathrm{H1}}, \Dh_{\mathrm{L1}}, \ldots\right\}\)
to be the labelling of a clock, \(P\left(\left\{\Dh_{\mathrm{H1}},
\Dh_{\mathrm{L1}}, \ldots\right\}\right) = 1 / \text{livetime}\).  In the
following subsections we show how each of the remaining terms in
\eqref{eqn2} and \eqref{eqn3} is obtained.

\subsection{\(P\left(\left\{\mathrm{H1}, \mathrm{L1}, \ldots\right\} |
\left\{\Dh_{\mathrm{H1}}, \Dh_{\mathrm{L1}}, \ldots\right\}, \signal
\right)\)}
\label{section1}

To obtain the probability that a signal yields coincident above-threshold
events in exactly some combination of instruments, we begin by assuming
that the coincidence time window is sufficiently large that if a signal is
seen above the \ac{SNR} threshold in two instruments the threshold
crossings will occur within that pair's coincidence window with certainty.
This assumption is essentially equivalent to requiring the coincidence
window for each pair of instruments to be at least as large as the sum of
their reciprocal bandwidths plus the light travel time between them (at
most few tens of milliseconds for ground-based laser interferometer
antennas).  If a more sophisticated ranking statistic is considered in the
future in which the time-of-arrival of the events in each of the
instruments is treated more carefully than to simply construct a pass/fail
test, or if the coincidence window for the pass/fail test is small, then
this assumption would need to be revisited.

Given our assumption about the formation of coincidences, we need only
consider the probability that the signal is seen above the \ac{SNR}
threshold.  The probability that a detectable signal is visible in exactly
some collection of instruments, \(\left\{\mathrm{H1}, \mathrm{L1},
\ldots\right\}\), (and no others) can be obtained to satisfactory accuracy
by Monte Carlo integration.  Referring to Appendix \ref{appendix2}, a
uniformly-distributed source position is selected on the sky and the
antenna responses to the \(+\) and \(\times\) polarizations computed, the
cosine of the orbit inclination is drawn uniformly from \([-1, +1]\), and
from the antenna responses and orbit inclination the quantity \(\Dnl\) is
computed for each instrument in the network using \eqref{eqn6}.  The number
of sources visible above the \ac{SNR} threshold \(\snr^{*}\) in two
instruments whose sensitivities are characterized by \(\Dnl_{1}\) and
\(\Dnl_{2}\) is
\begin{equation}
\label{eqn14}
\propto \int_{0}^{\infty} Q_{1}\left(\frac{\Dnl_{1}}{D}, \snr^{*}\right)
Q_{1}\left(\frac{\Dnl_{2}}{D}, \snr^{*}\right) D^{2} \diff D,
\end{equation}
while the number of sources visible to two instruments and not a third is
\begin{equation}
\label{eqn8}
\propto \int_{0}^{\infty} Q_{1}(\frac{\Dnl_{1}}{D}, \snr^{*})
Q_{1}(\frac{\Dnl_{2}}{D}, \snr^{*}) \left[ 1 - 
Q_{1}(\frac{\Dnl_{3}}{D}, \snr^{*}) \right] D^{2} \diff D,
\end{equation}
and so on.  All of these integrals diverge for the same reason that
\eqref{eqn7} diverges:  \(Q_{1} \neq 0\) as \(D \rightarrow \infty\).  The
probabilities that we seek will be the ratios of sums of integrals like
\eqref{eqn14} and \eqref{eqn8}, and even though the integrals diverge it
might be the case that the ratios we seek are well-defined;  or it might be
that the ratios can be made to be well-defined by introducing a
regularization, possibly one derived from the constraint that we seek a
ranking statistic to extremize detection efficiency at fixed \ac{FAR}.

\begin{figure}
\resizebox{\linewidth}{!}{\includegraphics{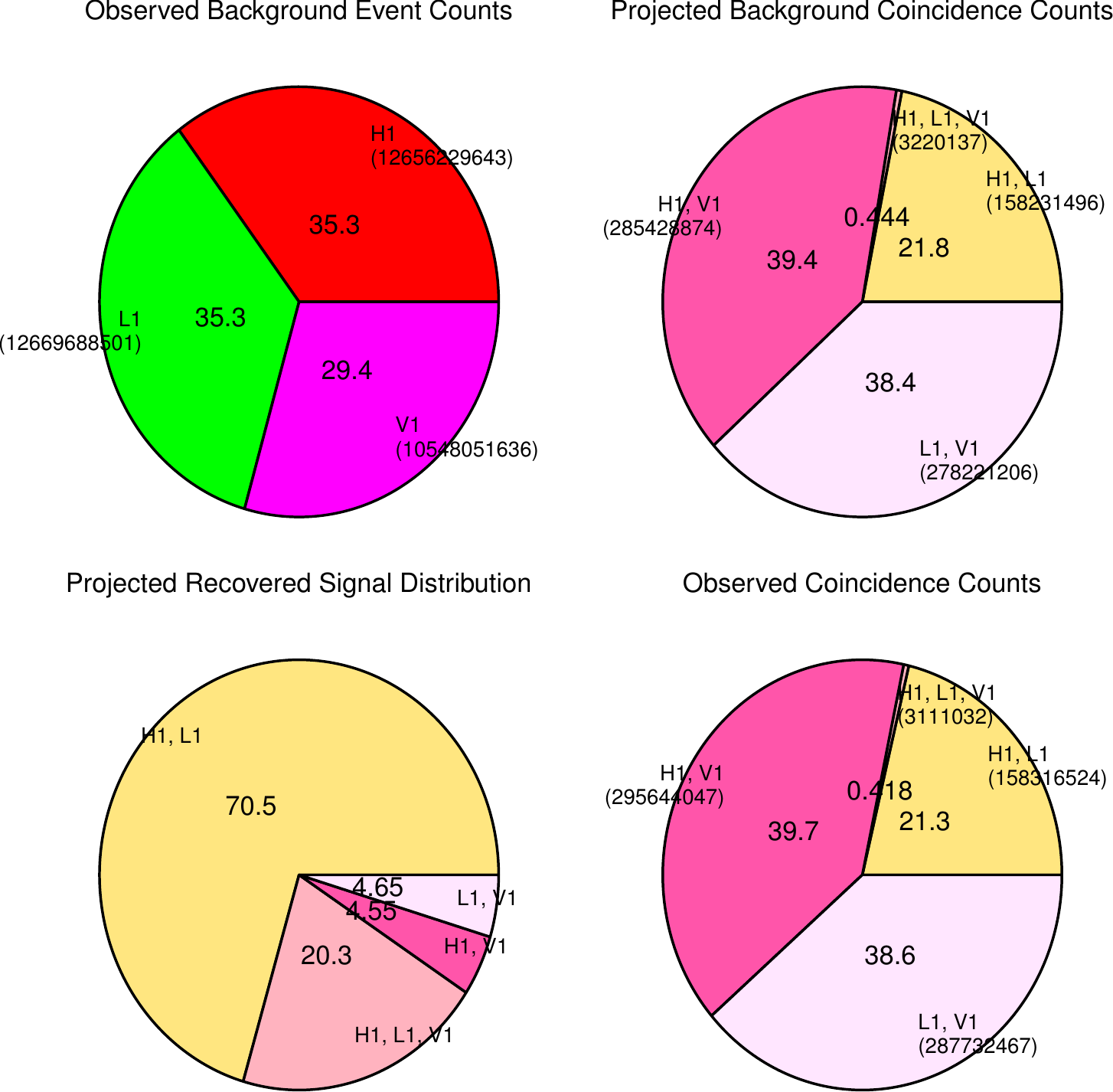}}
\caption{(Top Left) Total number (in parentheses) of non-coincident,
``noise'', events observed in each instrument and percent of total.  (Top
Right) Total number (in parentheses) of noise coincidence events predicted
to be observed for each combination of participating instruments, and the
percent of total.  This is \(P\left(\left\{\mathrm{H1}, \mathrm{L1},
\ldots\right\} | \noise\right)\), see Section \ref{section2}.  (Bottom
Right) Total number (in parentheses) of coincidence events observed for
each combination of participating instruments, and the percentage of the
total.  Notice the excellent agreement with the predicted percentages.
(Bottom Left) Probability that a signal visible to at least two instruments
is visible to the given combinations of instruments.  This is
\(P\left(\left\{\mathrm{H1}, \mathrm{L1}, \ldots\right\} |
\left\{\Dh_{\mathrm{H1}}, \Dh_{\mathrm{L1}}, \ldots\right\}, \signal
\right)\), see Section \ref{section1}.}
\label{fig1}
\end{figure}

Unfortunately we have not been able to make progress in this direction.
Instead, we proceed as follows.  We take \eqref{eqn9} to give the rate (up
to an irrelevant constant) at which signals in the given direction and with
the given polarization and given orbit inclination are seen by each
instrument in the network, and then assume that all signals visible to the
least sensitive instrument are visible to the next least sensitive
instrument with certainty, and so on.  This works best if the instruments
are not all very similar in sensitivity, which is likely to be the case as
the network of ground-based \ac{GW} antennas is commissioned in a staggered
schedule.  With that assumption, from differences and ratios of the rates
one can compute the probability that a source will be visible to some
combination of instruments (and no others) assuming that it is visible to
at least two.  These probabilities are added to a histogram and the process
of selecting a sky location and so on is repeated.  When sufficient
iterations have been made, the histogram is normalized.  We find that after
500,000 iterations the variance of the result is reduced to one part in
\(\order(10^{6})\), and a Python implementation of the sampling loop using
antenna response code implemented in C in LALSuite \cite{LALSuite}
completes in about \(\unit{30}{\second}\) on typical modern hardware.  An
example for \(\snr^{*} = 4\) is shown in the bottom-left panel of
\figref{fig1}.

\subsection{\(P\left(\snr_{\mathrm{H1}}, \snr_{\mathrm{L1}}, \ldots |
\left\{\Dh_{\mathrm{H1}}, \Dh_{\mathrm{L1}}, \ldots\right\},
\left\{\mathrm{H1}, \mathrm{L1}, \ldots\right\}, \signal\right)\)}

The calculation of the joint \ac{PDF} of the \acp{SNR} involves the same
framework as was used in Section \ref{section1}, and encounters some of the
same difficulties.  The \acp{PDF} are computed using Monte Carlo
integration of the density at each site in a multi-dimensional \ac{SNR}
histogram.  Each axis is binned using an \(\tan^{-1} \ln\) binning as
described in Appendix \ref{appendix1}.  Referring to Appendix
\ref{appendix2}, a uniformly-distributed direction is selected on the sky
and the antenna responses to the \(+\) and \(\times\) polarizations
computed, the cosine of an orbit inclination is drawn uniformly from \([-1,
+1]\), and from these the quantity \(\Dnl\) is computed for each instrument
in the network using \eqref{eqn6}.

\begin{figure}
\resizebox{.6\linewidth}{!}{\includegraphics{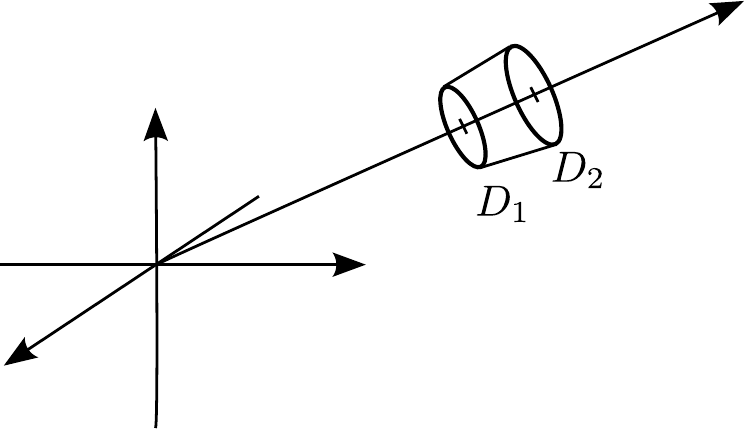}}
\caption{The number of sources in the plug of space is proportional to its
volume, which is \(\propto D_{2}^{3} - D_{1}^{3}\).}
\label{fig2}
\end{figure}

We identify the most sensitive of the instruments that must see the source,
and step through a sequence of bins of nominal \ac{SNR}, \(\snr_{0}\), in
that instrument.  Each bin has a lower bound, an upper bound, and a central
value.  From the \(\Dnl\)'s and each bin's central value, the \(\snr_{0}\)
in all other instruments are computed.  With respect to the instruments
that must not see the signal, we make the assumption that if the nominal
\ac{SNR} in those instruments is below the detection threshold,
\(\snr^{*}\), then they are all blind to it with certainty, and if above
\(\snr^{*}\) in one of them it sees it with certainty.  We are
approximating the Marcum Q-function with a Heaviside function for
performance purposes, and will rely on the superposition of many trials
combined with a convolution of the histogram with a density estimation
kernel to hide the approximation.  The binning of \(\snr_{0}\) in the most
sensitive instrument is started at 1, and terminated when this latter
condition is met.  Using \eqref{eqn6} and \(\Dnl\) in the most sensitive
instrument, the lower and upper bounds of each \(\snr_{0}\) bin can be
converted into upper and lower bounds, respectively, on the physical
distances of the corresponding sources.  The number of sources in the bin
is proportional to the difference of the cubes of these, see \figref{fig2}.
The \ac{SNR} \ac{PDF} is computed by iterating over the \(\snr_{0}\)
binning, at each bin computing a vector of Rician-distributed \acp{RV}
having noncentrality parameters given by the bin centre's co-ordinates to
give a vector of observed \acp{SNR}, and adding \(D_{2}^{3} - D_{1}^{3}\)
to that location in the binned \ac{PDF}.  A new sky location, etc., is
chosen and the whole process is repeated.  Note that this procedure results
in each bin in the \ac{SNR} \ac{PDF} containing a number proportional to
the rate of signals in that bin, \emph{not} the \ac{PDF} in that bin.

\begin{figure}
\resizebox{\linewidth}{!}{\includegraphics{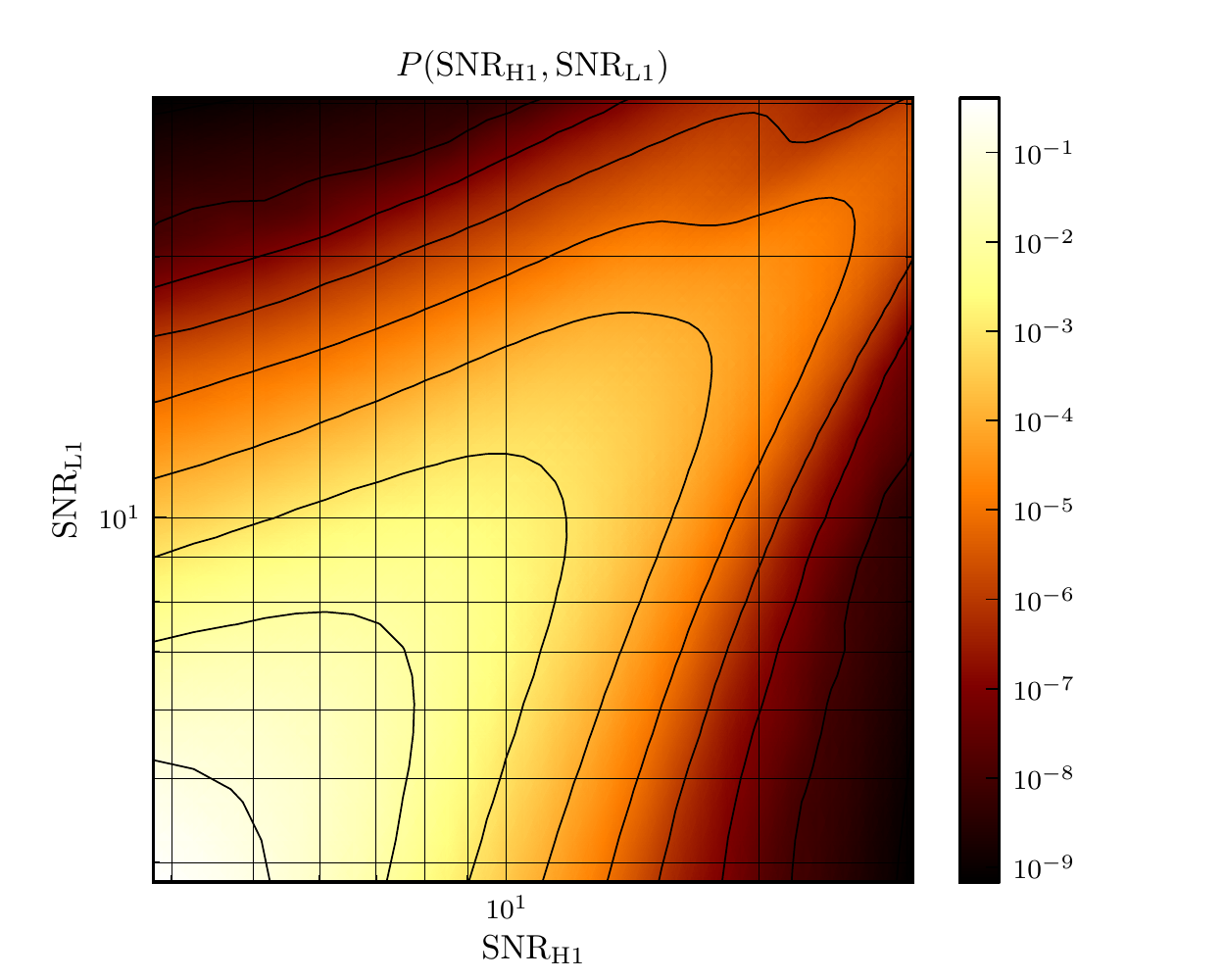}}
\resizebox{\linewidth}{!}{\includegraphics{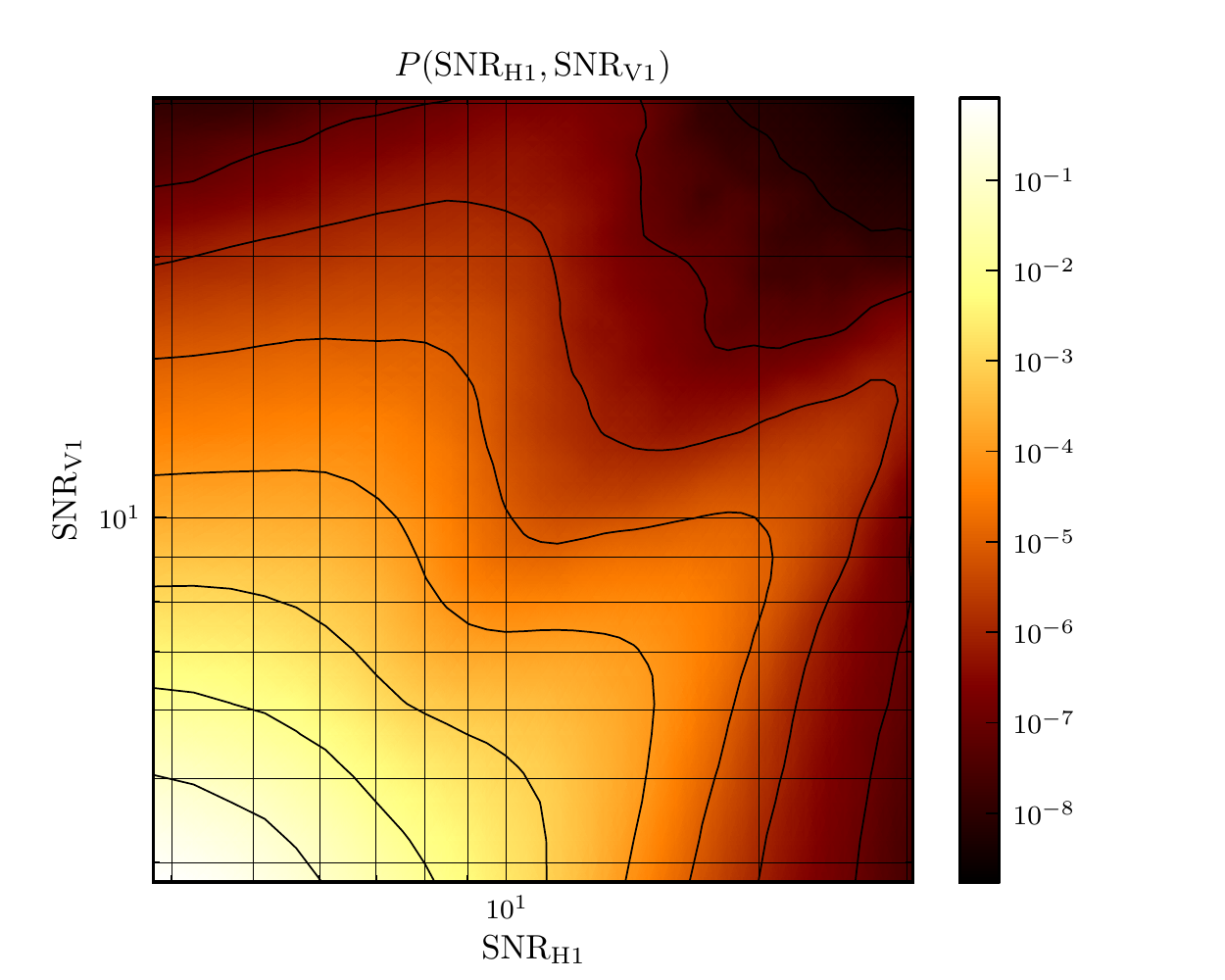}}
\resizebox{\linewidth}{!}{\includegraphics{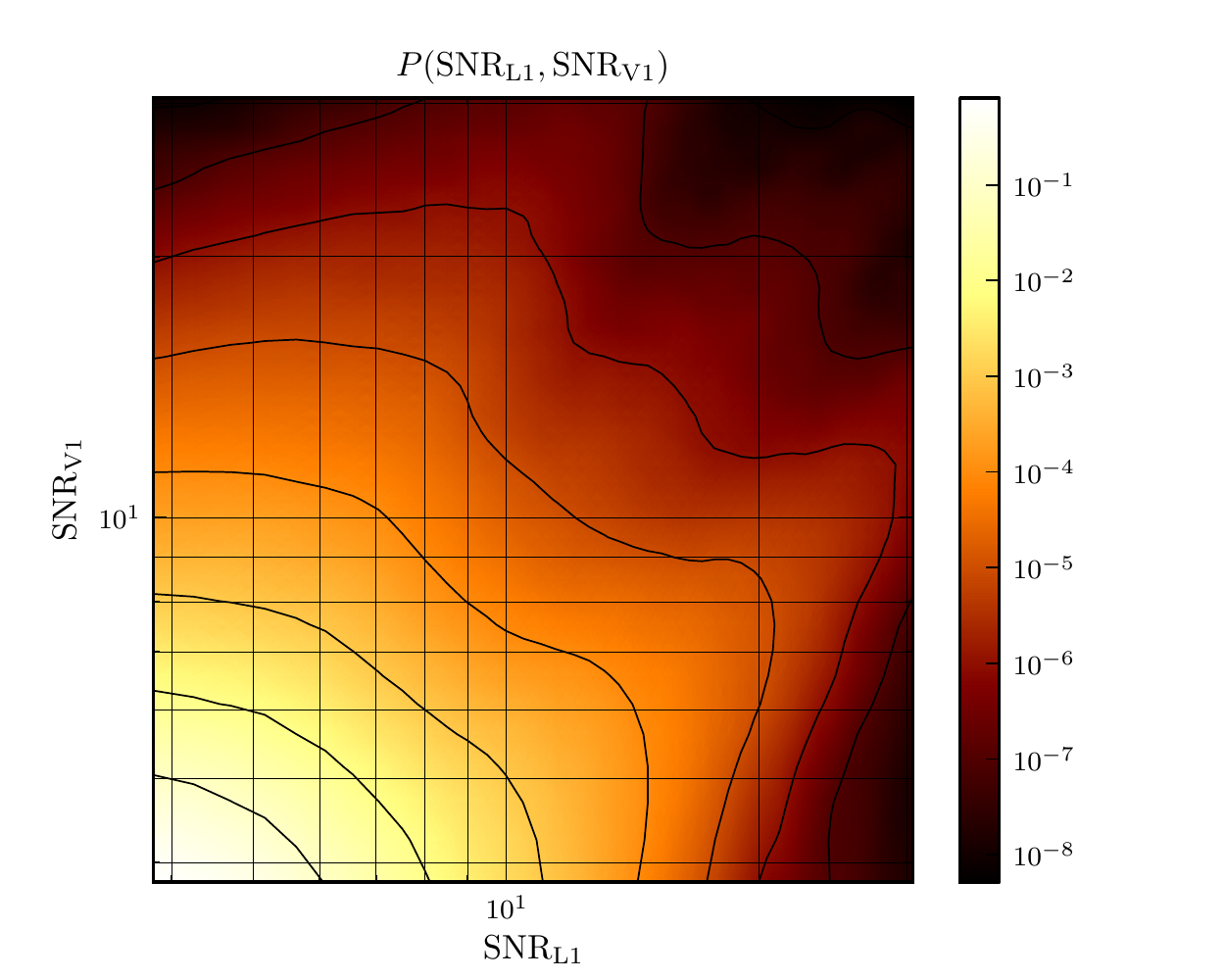}}
\caption{Examples of joint \acp{PDF} for \acp{SNR} in the three pairs of
instruments in the test analysis.}
\label{fig4}
\end{figure}

After a satisfactory number of iterations has been completed, the array of
signal rates is convolved with a Gaussian density estimation kernel, all
bins below \(\snr^{*}\) in any instrument are zeroed, and the array divided
by its sum (so that its sum is 1).  Finally each bin is divided by its
volume in \ac{SNR} to yield the properly-normalized \ac{PDF}.  Examples of
\acp{PDF} obtained this way are shown in
\figref{fig4}.
For these, the \(\tan^{-1} \ln\) binnings along each axis are given by
\(x_{\mathrm{lo}} = 3.6\), \(x_{\mathrm{hi}} = 120\), \(n = 100\), 80,000
iterations of the outer sky-location loop were performed, and a Gaussian
density estimation kernel with a standard deviation of 1.875 bins was
employed.

\subsection{\(P\left(\chisq | \snr, \signal\right)\)}

We begin by obtaining \(P\left(\snr, \chisq / \snr^{2} | \signal\right)\)
as described in \cite{cannon2012b}, using an analytic expression obtained
by assuming recoverable signals are to be found in glitch-free, Gaussian,
noise.  As before, the implementation samples the \ac{PDF} on a discrete
rectangular grid of \(\chisq / \snr^{2}\) vs.\ \(\snr\) in order to better
fit the natural shapes of the \ac{PDF}'s contours to the rectangular grid
of bins.  For both the \(\snr\) and \(\chisq / \snr^{2}\) axes we now use
the \(\tan^{-1}\ln\) binnings described in Appendix \ref{appendix1}.  The
\(\snr\) axes are binned using \(x_{\mathrm{lo}} = 3.6\), \(x_{\mathrm{hi}}
= 70\), \(n = 260\);  the \(\chisq / \snr^{2}\) axes using
\(x_{\mathrm{lo}} = .001\), \(x_{\mathrm{hi}} = .5\), \(n = 200\).

Once that \ac{PDF} is obtained, columns of constant \(\snr\) are
renormalized so that their integrals are 1, converting the function
represented by the array from a two-dimensional density to a 1-dimensional
density in \(\chisq / \snr^{2}\) parameterized by \(\snr\).

\subsection{\(P\left(\left\{\mathrm{H1}, \mathrm{L1}, \ldots\right\} |
\noise\right)\)}
\label{section2}

The search algorithm employs a bank of two-phase matched filters applied to
the strain time series recorded from each of the instruments.  When the
magnitude of the output of a filter crosses a preset threshold,
\(\snr^{*}\), a peak finder identifies the sample with the highest
\ac{SNR}, and that point in that filter's output is recorded as an event,
(sometimes called a ``trigger'', borrowing language from particle
experiments).  A ``coincidence'' occurs when two or more instruments
register events from the same filter within some window of time, \(|t_{1} -
t_{2}| \leq \tau_{12} = \delta t + \abs{\vec{x}_{1} - \vec{x}_{2}} / c\),
where \(\vec{x}_{1}\) and \(\vec{x}_{2}\) are the positions of the two
antennas and \(c\) is the speed of light.  When more than two instruments
participate in the coincidence all time differences between pairs of events
in the coincidence must satisfy the respective constraints.

There is a maximum rate at which events can be recorded (for example,
limited by the sample rate of the time series), but this limit is much
higher than the mean rate and so in the absence of signals the number of
events recorded in some interval from one filter is adequately approximated
as a Poisson process.  After measuring the mean rate of events in a filter
in each instrument, say by counting the number collected over some large
interval of time, it is straightforward to use the single-instrument rates
and the coincidence windows to compute the rates at which coincidences will
occur between pairs of instruments.  For example, if two instruments yield
events at mean rates \(\mu_{1}\) and \(\mu_{2}\), and the coincidence
window is \(\tau_{12}\), the mean rate at which coincidences occur is
\begin{equation}
\label{eqn10}
\mu_{1 \wedge 2}
   = 2 \mu_{1} \mu_{2} \tau_{12},
\end{equation}
because \(\mu_{1}\) is the number of events in instrument 1 per unit time,
and \(\mu_{2} \times 2 \tau_{12}\) is the number of events in instrument 2
one expects to find within \(\pm\tau_{12}\) of each of them.

Computing the rate at which higher-order coincidences occur is non-trivial,
and we do so using a stone-throwing technique.  For example, to compute the
rate at which triple coincidences occur, we start by computing
\begin{equation}
\label{eqn11}
\mu_{1 \wedge 2, 1 \wedge 3}
   = 2^{2} \mu_{1} \mu_{2} \mu_{3} \tau_{12} \tau_{13},
\end{equation}
giving the rate at which events in instrument 1 are found in coincidence
with events in both instruments 2 and 3.  The rate at which 3-way mutual
coincidences occur is
\begin{equation}
\mu_{1 \wedge 2 \wedge 3}
   = \mu_{1 \wedge 2, 1 \wedge 3} P(\text{2 and 3 are coincident} | 1),
\end{equation}
where \(P(\text{2 and 3 are coincident} | 1)\) is the probability that
events in instruments 2 and 3 are found to be in coincidence given that
they are both known to be coincident with an event in instrument 1.  This
probability is computed by stone throwing:  uniformly-distributed time
differences are chosen for instruments 1 and 2 and for instruments 1 and 3,
both consistent with the coincidence criteria for those pairs;  the time
difference between 2 and 3 is computed from these and if it satisfies the
coincidence criterion for that pair a successful outcome is recorded;  the
ratio of successful outcomes, \(n\), to total trials, \(m\), converges to
the desired probability as the number of iterations increases.  The number
of successful outcomes is a binomially-distributed \ac{RV} with standard
deviation \(\sqrt{m p (1 - p)} \leq \sqrt{m} / 2\) where \(p\) is the
probability of success \cite[Section 4.3]{winklerhays}.  We iterate until
the ratio of the upper bound of the standard deviation of the number of
successes to the observed number of successes, \(n\), falls below some
tolerance
\begin{equation}
\frac{\sqrt{m}}{2 n}
   < \epsilon.
\end{equation}
This stopping criterion requires a minimum of \(\epsilon^{-2} / 4\)
iterations before it can be met.  The number of iterations required, beyond
that, is minimized by choosing ``instrument 1'' to be the instrument for
which the Cartesian product of the coincidence windows between it and the
other instruments is smallest, so that coincidence with that instrument
provides the tightest prior constraint on the time differences among the
other instruments, and the rate of successful outcomes is maximized.

When some combination of instruments yields mutually coincident events, it
is possible those events are also coincident with an event in some other
instrument.  The rate at which coincidences occur that involve some
combination of instruments and no others can be found by subtracting from
that instrument combination's coincidence rate the rates of coincidences of
all proper supersets of that combination.  For example, if the network
consists of three instruments, the rate at which exactly instruments 1 and
2 are coincident is
\begin{equation}
\mu_{\ngtr 1 \wedge 2}
   = \mu_{1 \wedge 2} - \mu_{\ngtr 1 \wedge 2 \wedge 3},
\end{equation}
where \(\mu_{\ngtr 1 \wedge 2 \wedge 3} = \mu_{1 \wedge 2 \wedge 3}\)
because there are no other instruments.  Having obtained, for each
instrument combination, the rate at which coincidences occur involving
those instruments and no others, the probability that a coincidence chosen
at random involves exactly some combination of instruments is obtained by
the ratio of each such rate to the sum.  For example, in the
three-instrument case,
\begin{equation}
\label{eqn13}
P\left(\left\{ 1, 2 \right\} | \noise\right)
   = \frac{\mu_{\ngtr 1 \wedge 2}}{\mu_{\ngtr 1 \wedge 2} + \mu_{\ngtr 1
   \wedge 3} + \mu_{\ngtr 2 \wedge 3} + \mu_{\ngtr 1 \wedge 2 \wedge 3}}.
\end{equation}
An example of the result with \(\epsilon = 10^{-4}\), and a comparison of
the predicted to the observed behaviour is shown in \figref{fig1}.

\subsection{\(P\left(\snr_{\mathrm{inst}}, \chisq_{\mathrm{inst}} |
\noise\right)\)}

This \ac{PDF} is obtained exactly as described in \cite{cannon2012b}, by
histograming the properties of single-instrument events that fail to be
coincident with events in any other instruments.  As before, the
implementation uses a histogram binned on a rectangular grid in \(\chisq /
\snr^{2}\) vs.\ \(\snr\) in order to better fit the natural shapes of the
\ac{PDF}'s contours to the rectangular grid of bins.  For both the \(\snr\)
and \(\chisq / \snr^{2}\) axes we now use the \(\tan^{-1}\ln\) binnings
described in Appendix \ref{appendix1}.

\subsection{\(\lr\left(\sigparams\right) = P\left(\sigparams |
\signal\right) / P\left(\sigparams | \noise\right)\)}

In considering the Bayes factor for the intrinsic parameters of a
candidate, \(\sigparams\), we first point out that in the example
implementation, a ``coincidence'' is formed when two or more instruments in
the network register events from the \emph{same} filter within some window
of time.  Therefore, there is no distinction between the \(\sigparams\)
carried by single-instrument events and the \(\sigparams\) carried by a
coincidence.  Searches for \acp{GW} using algorithms that permit a waveform
mismatch between instruments in a coincidence will require a different
technique for computing the Bayes factor for \(\sigparams\) than presented
here.

The denominator, \(P\left(\sigparams | \noise\right)\), is obtained as a
by-product of the procedure in Section \ref{section2} for obtaining
\(P\left(\left\{\mathrm{H1}, \mathrm{L1}, \ldots\right\} | \sigparams,
\noise\right)\).  Recall that for brevity we have been omitting
\(\sigparams\) from the parameters of the \acp{PDF}, and that \eqref{eqn13}
was implicitly for a choice of \(\sigparams\).  The denominator in
\eqref{eqn13} gives the total rate of background coincident events expected
for a choice of \(\sigparams\).  \(P\left(\sigparams | \noise\right)\) is
obtained by dividing that rate by the sum of those rates for all
\(\sigparams\).

\(P\left(\sigparams | \signal\right)\) is chosen by the individuals
performing the search.  Specifying the relative frequency at which
templates are expected to yield candidates as a result of genuine signals
is how the ranking statistic is informed of ones prior belief about the
distribution of the intrinsic parameters carried by signals of
astrophysical origin.  In the example implementation demonstrated here,
\(P\left(\sigparams | \signal\right)\) was chosen to be uniform on the
manifold in waveform space comprising the template bank, \textit{i.e.}, all
templates are considered equally likely in the signal population.  This
prior is the one implicitly chosen in past searches for \acp{GW} from
\acp{CBC}.  Because the waveform space's metric is determined by the noise
spectrum of the antennas, \textit{i.e.}, not by the physical properties of
the sources, this prior is not uniform in any astrophysically meaningful
quantities, like the mass of the source.  Other choices might be better in
future searches.  One could conceivably fix \(P\left(\sigparams |
\signal\right)\) to be uniform in template index, and vary the template
density to affect the desired prior in astrophysical parameters.  This is
the most computationally efficient as it avoids, maximally, the filtering
of templates whose events are subsequently discarded by the ranking
statistic, but one needs to be certain of the choice of prior in advance
because the \ac{SNR} lost in low-density regions of the template bank
cannot be recovered without re-analyzing the data if one decides that a
different prior is desired.  Although searches for \acp{CBC} have not done
so, template-based searches for cosmic string bursts
\cite{s56vsr123strings} and for \acp{CW} from pulsars \cite{jaranowski1998}
do incorporate astrophysical priors into their template weighting and/or
placement.

\subsection{Time Dependence}

As discussed above, we have constructed a framework that allows us to
include the time dependence of the antenna noise floors (encoded via
horizon distance) in the ranking statistic.  The daily cycle of human
activity, weather, and thermal changes to optics can alter the Gaussian
noise floor of the antennas substantially on time scales of hours, these
changes are easily tracked on time scales of minutes, and so there is both
a desire and an opportunity to incorporate knowledge of Gaussian noise
floor fluctuations in the assessment of candidates on an event-by-event
basis.

Other quantities change in time as well.  The mean background event rates
change, and the antennas undergo periodic maintenance that can alter the
statistical properties of their noise processes, for example changing the
distribution of \(\chisq\) parameters.  We do not explicitly include the
time dependence of these other quantities in the description of the ranking
statistic given here.  In practice it is found that a week or more of data
is required before the histograms used to model the noise \acp{PDF} in the
ranking statistic's denominator have converged enough that the ranking
statistic value assigned to a given candidate is, effectively, stable.
Therefore it is not possible to monitor or even define changes to these
\acp{PDF} on time scales less than \(\unit{\order(1)}{week}\).

At the same time, to simplify the management of the computer systems, one
tends to analyze data in blocks of time, and these blocks tend to be of
about the same duration as is required to stabilize the denominator
\acp{PDF}.  Therefore, through this piecewise analysis of the data, by
simply collecting fresh histograms within each block of time the time
dependence of, for example, the \(\snr, \chisq\) histograms, is naturally
incorporated into the ranking statistic.  Finally, as was done in Section
\ref{section2} to normalize the ranking statistic across different
instrument combinations, recording the mean background event rate in each
interval of time allows the ranking statistic to be normalized across
disjoint analysis blocks providing a single, unified, ranking statistic for
the entire experiment.

\section{False-Alarm Probability}

We now have all the ingredients of the ranking statistic, and so at the end
of a search for \acp{GW} from \acp{CBC} we can order the candidate events
from most signal-like to least signal-like.  We wish to know how
significant a discovery the most signal-like of the candidates are:  should
we have expected to see things so signal-like by random chance from the
noise, or might these be genuine signals?  To accomplish this, we require
\begin{equation}
\label{eqn12}
P(\lr | \sigparams, \noise)
   = \int_{\Sigma(\lr)} P(\ldots | \sigparams, \noise) \diff^{n - 1}
   \Sigma,
\end{equation}
the \ac{PDF} for \(\lr\) (for a choice of intrinsic parameters) in the
noise population.  This is obtained by integrating the noise \ac{PDF} for
all \(n\) parameters in the ranking statistic over \(n - 1\) dimensional
surfaces, \(\Sigma\), of constant \(\lr\).  The \ac{CCDF} of \eqref{eqn12}
gives the probability that a noise event drawn at random is found to have a
value of \(\lr\) at least as high as some threshold, and from that one can
obtain the probability that a signal-free experiment of some duration (that
yields some number of events) yields one or more events with values of
\(\lr\) at least as high as some threshold --- the \ac{FAP}.  The details
of this were discussed in \cite{cannon2012b}, here we will describe how
\eqref{eqn12} is evaluated in the context of the new expression for
\(\lr\).

We do not know the surfaces of constant \(\lr\), instead we construct an
approximation of \(P(\lr | \sigparams, \noise)\) by visiting points in the
\(n\)-dimensional parameter space, evaluating \(\lr\) and \(P(\ldots |
\sigparams, \noise)\) at each point that we visit (``\(\ldots\)'' are the
\(n\)-dimensional co-ordinates of the point), and constructing a histogram
of \(\lr\) with each sample weighted by \(P(\ldots | \sigparams, \noise)\).
As explained in Section \ref{section3}, in \cite{cannon2012b} the
approximation of the \ac{LR} as a product of several low-dimensional terms
allowed this sampling procedure to be performed by exhaustively visiting
every point in the discretely-sampled \acp{PDF}.  Because, here, the
numerator of \(\lr\) no longer factors into independent single-instrument
terms we can no longer use this technique, exactly.  Instead of factoring
\(\lr\) and exhaustively evaluating it at each grid point in the
single-instrument \acp{PDF}, we leave it as an \(n\)-dimensional function
and evaluate it at a randomly-selected subset of grid points in the
\(n\)-dimensional space, and histogram those results alone.  As the number
of samples gets large the histogram of \(\lr\) samples converges to \(P(\lr
| \sigparams, \noise)\).  The rate of convergence can be increased by
adjusting the \ac{PDF} from which grid points are drawn (and re-weighting
the samples appropriately).  This technique for approximating integrals is
known as importance-weighted sampling \cite[Section 11.7.2]{landaupaez}.

\begin{figure}
\resizebox{\linewidth}{!}{\includegraphics{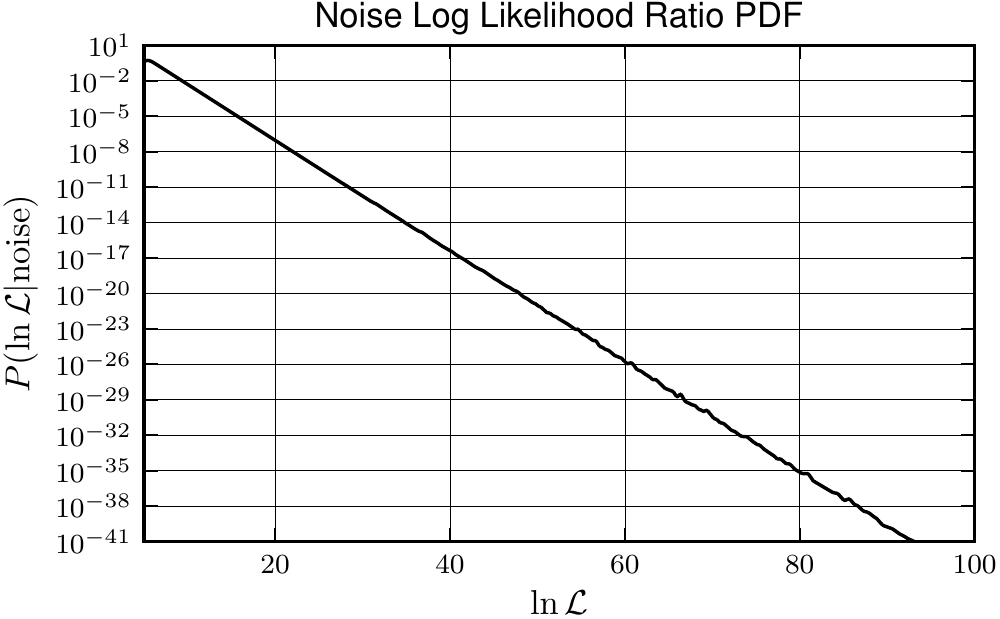}}
\resizebox{\linewidth}{!}{\includegraphics{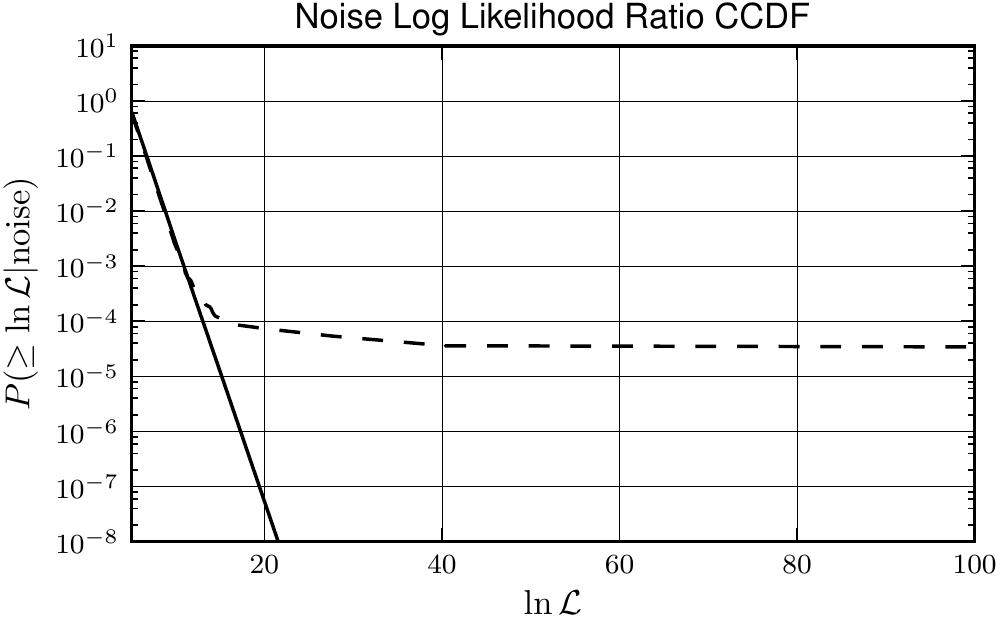}}
\caption{(Top) \ac{PDF} for \(\ln \lr\) obtained from the
importance-weighted sampling integration of the noise model.  This is, in
effect, a normalized histogram of the samples constructed with the
\(\tan^{-1} \ln\) binning described in the text.  Note the enormous dynamic
range on both axes that has been captured by the binning and
importance-weighted sampling.  (Bottom) \ac{CCDF} for the noise model
(solid) obtained by integrating the \ac{PDF} in the top panel, shown
together with the actual observed \ac{CCDF} (dashed).  The departure of the
observed \ac{CCDF} from the model is due to an astrophysically realistic
population of injections that was present in this data set, but note the
excellent agreement in the bulk.}
\label{fig3}
\end{figure}

The total number of distinct \(n\)-dimensional grid points in the binning
used to represent \(\lr\) is enormous --- in our example it's over \(5.6
\times 10^{14}\) --- but we find that a satisfactory approximation of the
\ac{PDF} can be obtained after just \(4 \times 10^{7}\) samples.  The
result is shown in \figref{fig3}.
This was obtained using an \(\tan^{-1}\) binning for \(\ln \lr\) as
described in Appendix \ref{appendix1} with \(\ln x_{\mathrm{lo}} = 0\),
\(\ln x_{\mathrm{hi}} = 110\), and \(n = 3,000\).  All indexes for the
binnings comprising the space over which \(\lr\) is defined were drawn from
uniform distributions except the \(\snr\) bins whose indexes were drawn
from a distribution whose \ac{CDF} is proportional to \((\text{bin
index})^{0.8 / (\text{\# instruments})^{3}}\) and restricted to bins not
below the single-instrument \ac{SNR} threshold, \(\snr^{*}\).  This
distribution was arrived at empirically to efficiently favour interesting
\acp{SNR}.  The counts in the \(\ln \lr\) histogram were convolved with a
Gaussian density estimation kernel having a standard deviation of 8 bins,
the array of counts then rescaled so that its sum was 1, and then each bin
divided by its size in \(\ln \lr\) to yield a normalized,
discretely-sampled, \ac{PDF}.

Note that the \ac{PDF} and \ac{CCDF} have been truncated at \(\ln \lr =
5\).  The procedure by which candidate events are collected employs a
clustering step to remove redundant candidates resulting from each signal.
It has been found that above \(\ln \lr = 5\) the noise process yields
candidates at a sufficiently low rate that their population is unaffected
by the clustering.  Modelling the \ac{PDF} for the noise process below \(\ln
\lr = 5\) requires the clustering process to be taken into consideration,
and this will be the subject of future work.  The remainder of the
procedure for converting \(P(\lr | \sigparams, \noise)\) into a mapping
from \(\lr\) to \ac{FAP} and \ac{FAR} is exactly as in \cite{cannon2012b},
and without the clustering model is applicable only above the cut-off
threshold of \(\ln \lr = 5\).

\section{Event Rate Estimation}

\begin{figure}
\resizebox{\linewidth}{!}{\includegraphics{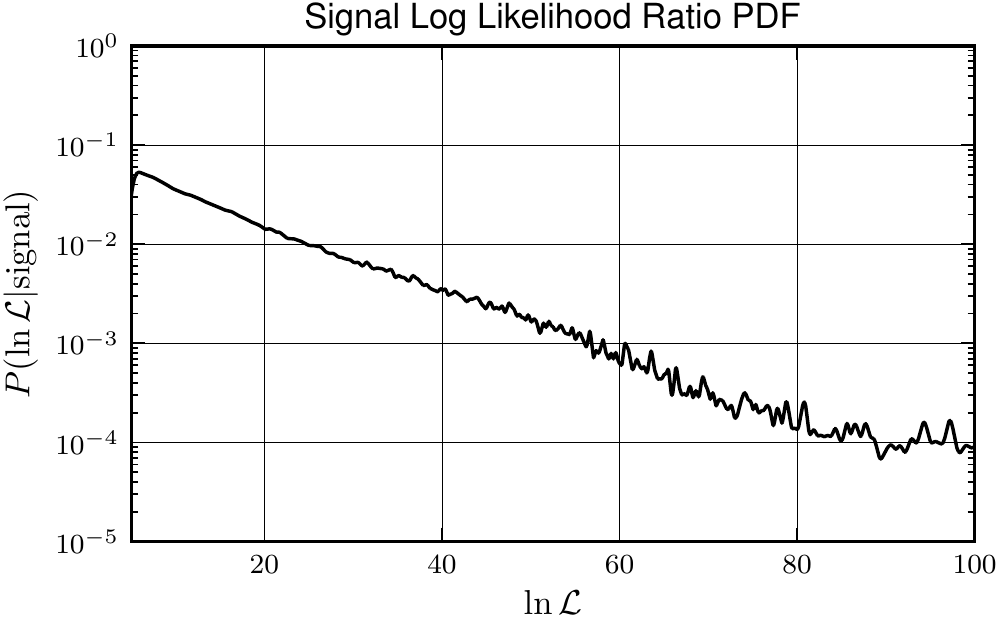}}
\resizebox{\linewidth}{!}{\includegraphics{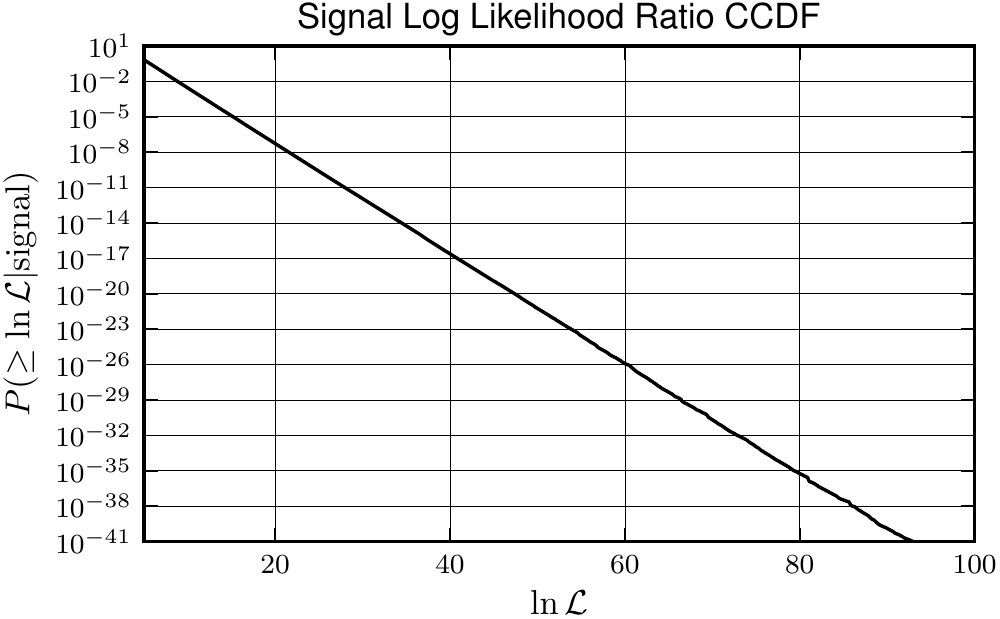}}
\caption{(Top) \ac{PDF} for \(\ln \lr\) obtained from the
importance-weighted sampling integration of the signal model.  (Bottom)
\ac{CCDF} for the signal model obtained by integrating the \ac{PDF} in the
top panel.  The fluctuations observed in the PDF at high values of the
ranking statistic appear to be due to sampling noise in the \ac{SNR}
\acp{PDF} at high \ac{SNR}, in particular the fluctuations are not the
result of under-sampling the ranking statistic:  running the stochastic
integration for more iterations, changing the random number generator's
seed, and changing the distribution from which samples are drawn in the
integral are not found to alter the appearance of the peaks and troughs in
this PDF, but they are changed by generating new \ac{SNR} \acp{PDF} and
increasing the number of iterations in the \ac{SNR} \ac{PDF} sampler
decreases their amplitude.}
\label{fig5}
\end{figure}

Since we have a reasonable model for \(P(\ldots | \signal)\) in the
numerator of \(\lr\), the procedure described above for approximating
\eqref{eqn12} can also be used to approximate \(P(\lr | \sigparams,
\signal)\) by replacing \(P(\ldots | \sigparams, \noise)\) in the integrand
with \(P(\ldots | \sigparams, \signal)\).  In fact, the \emph{exact} same
procedure is used, and approximations of both \(P(\lr | \sigparams,
\noise)\) and \(P(\lr | \sigparams, \signal)\) are obtained simultaneously
in the same sampling loop.  Following through with a marginalization over
\(\sigparams\) yields \(P(\lr |\signal)\).  An example is shown in
\figref{fig5}, where both the \ac{PDF} and \ac{CCDF} have been truncated at
\(\ln \lr = 5\) to match what was done with the noise model.

\(P(\lr | \signal)\) and \(P(\lr | \noise)\) are the ingredients required
to implement the signal rate estimation technique of Farr \textit{et al.}\
\cite{farr2013a}.  Farr \textit{et al.} derive the joint \ac{PDF} for the
rates of signal events and noise events from the ranking statistic values
assigned to all events collected in the experiment, \cite[equation
(21)]{farr2013a}.  Marginalizing this \ac{PDF} over background rate yields
the posterior \ac{PDF} for the rate of signals during the experiment.  Farr
\textit{et al.}\ imply the possibility of obtaining a closed-form
expression for this posterior, but for experiments with a large number of
candidate events we find that approach to be intractable.  We resort to
Markov chain Monte Carlo sampling from the joint \ac{PDF} for the
marginalization.  For this we use the \texttt{emcee} sampler by
Foreman-Mackey \textit{et al.}\ \cite{emcee}.

\begin{figure}
\resizebox{\linewidth}{!}{\includegraphics{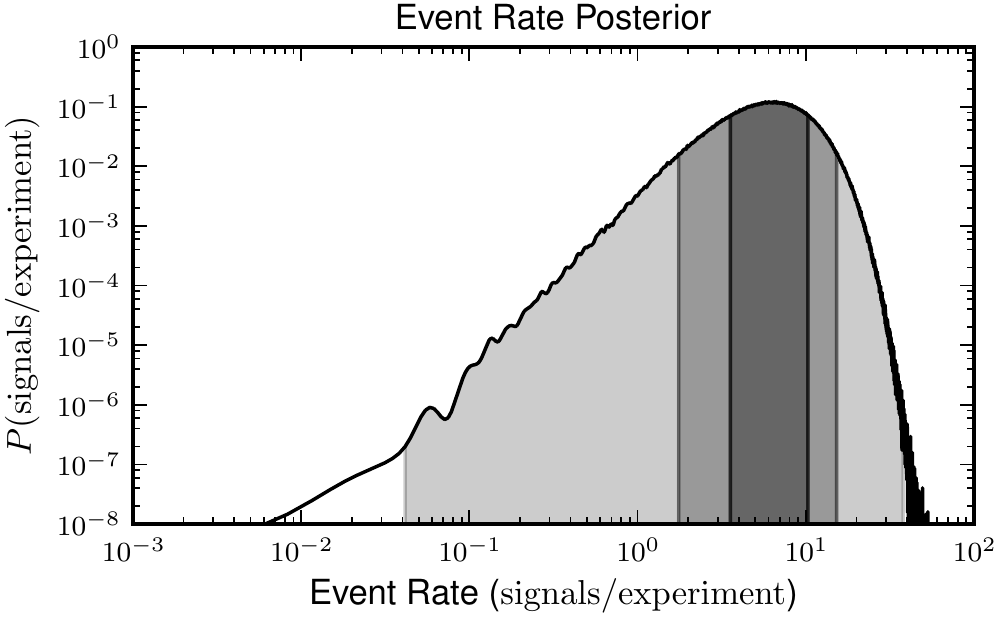}}
\caption{Signal rate posterior \ac{PDF}.  From darkest to lightest the
shaded areas indicate the 68\%, 95\% and 99.9999\% (``\(5 \sigma\)'')
credible intervals.  Note that the latter excludes 0.}
\label{fig6}
\end{figure}

We wish to obtain credible intervals from the signal rate posterior
\ac{PDF}.  In particular, we wish to know if the \(99.9999\%\) credible
interval excludes 0.  To achieve this, we need to measure the posterior's
tails well, and running the sampler long enough to do so using the correct
\ac{PDF} is inconvenient.  Therefore, we sample from the square root of the
joint \ac{PDF} given in \cite[equation (21)]{farr2013a}, and correct the
histogram of samples by squaring the count in each bin and dividing by the
bin size.
An example of the result is shown in \figref{fig6}.
This was obtained with 40 walkers, the chain burned-in for 1,000
iterations, then run for 400,000 iterations, and the samples histogrammed
using a uniform-in-logarithm binning spanning the range of values returned
from the sampler and having about 22,000 bins (so there is an average of
about 730 samples per bin).  Implementing the posterior function in C and
parallelizing the sum within it using OpenMP, the sampling process takes
several hours on a modern 16-core machine.  After correcting the bin counts
as described above, the corrected counts were convolved with a Gaussian
density estimation kernel having a standard deviation of 5 bins, the
convolved bin counts normalized to have a sum of 1, and finally divided by
the bin sizes to yield the normalized discretely-sampled estimate of the
posterior \ac{PDF}.

The mean of the signal rate posterior \ac{PDF} in the example is
\(\unit{8.0}{experiment^{-1}}\), the maximum likelihood signal rate is
\(\unit{6.7}{experiment^{-1}}\), the 68\% credible interval is
\(\unit{[3.6, 10]}{experiment^{-1}}\) , the 95\% credible interval
\(\unit{[1.8, 15]}{experiment^{-1}}\), and the 99.9999\% credible interval
is \(\unit{[.040, 38]}{experiment^{-1}}\) which, in particular, excludes 0.
In our example analysis, we have the benefit of knowing the list of
synthetic signals that were added to the data, and we can consult that list
to see how many of them are expected to be visible above the \ac{SNR}
threshold in at least two instruments.  That number as a function of the
\ac{SNR} threshold is shown in \figref{fig7}.  The example analysis used a
single-instrument \ac{SNR} threshold of \(\snr^{*} = 4\), and apparently
19.4 signals are expected to have yielded candidate events involving a
coincidence between at least two instruments.  Additionally, as discussed
above, to remove the need to model clustering survival, a ranking statistic
cut has been imposed discarding all events with \(\ln \lr < 5\).  Of the
19.4 signals expected to yield coincident events, only 11.5 are expected to
survive the ranking statistic cut.\footnote{This is arrived at by drawing
\ac{SNR} and \(\chisq\) values from distributions constructed for each
injection, evaluating the ranking statistic at those values, and measuring
the survival probability for each injection by iterating.  That analysis
indicates that 12.1 signals are expected to survive all cuts, but at the
ranking statistic threshold the clustering is already causing about 5\% of
events to be lost so we suppose that 0.6 more events are lost due to
clustering and conclude that about 11.5 should remain.}  This rate of
signals is within the 75\% credible interval, and so there is already
reasonable agreement between the posterior \ac{PDF} and the known expected
signal rate, but it is also possible that a bias is present.  In
considering the origin of a bias, our simple choice of \(P(\sigparams |
\signal) = \text{const}\) does not correspond to the probabilities with
which signals are expected to be recovered by different templates, and some
bias is to be expected as a result of this.  Probably the dominant reason
for a bias, however, is that the injected signals were drawn from a
population of sources with spinning components whereas the templates used
to identify candidates were non-spinning and therefore would not have
recovered as much \ac{SNR} as they could have, and even small losses can
significantly impact the expected rate.  For example, assuming the mismatch
due to spin results in an average loss of \ac{SNR} of just 6\% reduces the
expected signal recovery rate to \(\unit{8.9}{experiment^{-1}}\), which is
within the 50\% credible interval, and in essential agreement with the
posterior \ac{PDF}.  A detailed investigation of the \ac{SNR} recovery
efficiency of the template bank used in the example search is beyond the
scope of this work;  the naive assumption that the template bank efficiency
is given by the template bank's density already yields good agreement
between the posterior \ac{PDF} for the signal rate and the signal
population, and so if there is a bias in the signal rate posterior it is
not greater than can be explained by assuming a few percent loss of
\ac{SNR}.

\section{Conclusion}

We have shown the construction of a new ranking statistic for searches for
\acp{GW} from \acp{CBC}, and demonstrated its implementation including an
exposition of the technical details required to do so.  We have shown that
the ranking statistic continues to allow us to construct a mapping to
\ac{FAP} without the use of time slides.  We have shows that the ranking
statistic can now be used with the rate estimation technique of Farr
\textit{et al.}\ \cite{farr2013a} to obtain a reasonable posterior
\acp{PDF} for the rate of signals observed during an experiment.  We have
found that while the identification of single, statistically-significant,
outlier is not adversely affected by small \ac{SNR} losses arising from
signal/template mismatch, the inference of a signal rate from the
population of candidate events collected by the analysis can be sensitive
to small \ac{SNR} losses due to template bank inefficiency.  This is not
unexpected since the rate of detectable signals should scale cubically with
the \ac{SNR} collected by the templates, and we showed how adjustments to
the expected rate of recovered signals resulting from as little as a 6\%
\ac{SNR} loss in the template bank can bring the measured signal rate
posterior \ac{PDF} into good agreement with the population of signals we
know to be in the test data.

Future work will look more carefully at the problem of interpreting the
signal rate posterior \ac{PDF} in terms of an astrophysical merger rate.
In particular, we will examine the problem of modelling the loss of
low-significance events due to clustering in more detail, and incorporate a
more detailed analysis of the \ac{SNR} recovery efficiency of template bank
in the interpretation.

\acknowledgments

KC and JP were supported by the National Science and Engineering Research
Council, Canada.  This work has LIGO document number LIGO-P1400175.  We
thank the UW-Milwaukee Center for Gravitation and Cosmology and the LIGO
Laboratory for use of its computing facility to make this work possible
through NSF grants PHY-0923409 and PHY-0600953.  We thank the LIGO
Scientific Collaboration and the Virgo Scientific Collaboration for
providing us with the data to test the methods described in this work.
Figures were generated with Matplotlib \cite{Hunter:2007}.  KC thanks
Jolien Creighton for his careful reading of the manuscript and many helpful
insights that have improved the efficiency of the implementation of the
techniques described herein, and thanks Ilya Mandel for his assistance
improving this presentation of our results.  All remaining problems are our
own.

\appendix

\section{Distance and \ac{SNR}}
\label{appendix2}

For convenience, we collect here some relationships relating the distance
to a \ac{GW} source, its \ac{SNR}, and the number density of sources.

The strain seen in a \ac{GW} antenna is the projection of the \ac{GW} field
onto the antenna's response tensor, and in terms of the two transverse,
traceless, polarization components of the field, \(h_{+}\) and \(h_{\times}\),
can be written \cite[Section 9.4]{300yearschapter9}
\begin{equation}
h(t)
   = F_{+}(\phi, \theta, \psi) h_{+}(t) + F_{\times}(\phi, \theta. \psi)
   h_{\times}(t).
\end{equation}
where the antenna response factors, \(F_{+}\) and \(F_{\times}\), depend on
the direction to the source, \((\phi, \theta)\), and the orientation of the
polarization axes, \(\psi\) \cite[equation (B6)]{Anderson:2000yy}.  For
non-precessing \acp{CBC} whose radiation is dominated by the \(l, m = 2,
2\) mode, \(h(t)\) is inversely proportional to the effective distance
\cite[equation (3.3c)]{findchirppaper}
\begin{equation}
\label{eqn4}
\Deff
   = D \left[ F_{+}^{2} \left( \frac{1 + \cos^{2} \iota}{2} \right)^{2} +
   F_{\times}^{2} \cos^{2} \iota \right]^{-\frac{1}{2}},
\end{equation}
where \(D\) is the physical distance to the source, and \(\iota\) the angle
between the line-of-sight to the source and its orbital axis.  \(\Deff \geq
D\), and a source is ``optimally oriented'' with respect to an antenna when
\(\Deff = D\).

Defining \(\snr_{0}\) to be the nominal matched-filter output observed for
a signal in the absence of noise, \(\snr_{0} \propto h(t) \propto
\Deff^{-1}\) \cite{findchirppaper}.  This relationship can be written as
\begin{equation}
\label{eqn5}
\snr_{0}
   = 8 \Dh / \Deff,
\end{equation}
thereby defining \(\Dh\), the ``horizon distance'', the physical distance
at which an optimally-oriented source is seen with a nominal \ac{SNR} of 8
\cite{findchirppaper}.  Note that \(\Dh\) depends on the antenna noise
spectrum and the physics of the \ac{GW} source:  one may speak of an
horizon distance being associated with a source by assuming a canonical
antenna noise spectrum, or with an antenna by assuming a canonical source;
stronger emitters of \acp{GW} are said to have larger \(\Dh\), less
sensitive antennas smaller \(\Dh\).  Combining \eqref{eqn4} and
\eqref{eqn5},
\begin{equation}
\label{eqn6}
D \snr_{0}
   = 8 \Dh \left[ F_{+}^{2} \left( \frac{1 + \cos^{2} \iota}{2} \right)^{2}
   + F_{\times}^{2} \cos^{2} \iota \right]^{\frac{1}{2}}
   = \Dnl,
\end{equation}
where the parameter \(\Dnl\) is introduced for compactness:  a sort-of
``noise-limit distance'', the physical distance at which a source with the
given geometric arrangement with respect to a given antenna is seen at an
\ac{SNR} of 1 in that antenna.

In stationary Gaussian noise the \ac{SNR}, \(\snr\), at which a signal is
recovered is a 2 \ac{DOF} noncentral \(\chi\)-distributed \ac{RV} with
noncentrality parameter \(\snr_{0}\) \cite[Section IV]{findchirppaper}.
This is also known as the Rice distribution with \(\sigma = 1\) and
noncentrality parameter \(\snr_{0}\) \cite[equation
(3.10-11)]{rice1945}.\footnote{Reference is to Rice's original work;  he
does not call the distribution by that name.} The probability that a source
with nominal \ac{SNR} \(\snr_{0}\) is observed above some \ac{SNR}
threshold \(\snr^{*}\) is
\begin{equation}
P(\snr \geq \snr^{*} | \snr_{0})
   = Q_{1}(\snr_{0}, \snr^{*})
   = Q_{1}(\frac{\Dnl}{D}, \snr^{*}),
\end{equation}
where \(Q_{1}\) is the first-order Marcum Q function \cite{marcum1960}.
Assuming the number density of sources to be uniform in volume, it is
\(\propto D^{2} \diff D = -\frac{\Dnl^{3}}{\snr_{0}^{4}} \diff \snr_{0}\),
therefore the total number of sources in a given direction with a given
polarization and orbit inclination visible to an antenna with a given
horizon distance is
\begin{equation}
\label{eqn7}
\text{\# sources}
   \propto \Dnl^{3} \int_{0}^{\infty} Q_{1}\left(\snr_{0}, \snr^{*}\right)
   \snr_{0}^{-4} \diff \snr_{0}.
\end{equation}
Because \(Q_{1}(0, \snr^{*}) \neq 0\) the integral diverges.  Physically,
assuming a uniform source density to arbitrary distance yields an infinite
number of sources at infinite distance with zero nominal \ac{SNR} that,
nevertheless, due to noise fluctuations have a non-zero probability of
registering as events.  Because the universe has a finite age and stellar
evolution must be allowed to progress for a period of time after the birth
of the universe before compact object mergers can occur, the source density
must, really, fall to 0 at some distance, and if this is correctly
accounted for this integral must be finite --- Olbers' paradox revisited.

If the number density cutoff occurs at a distance \(D \gg \Dnl\) (the
antennas cannot see as far as the first sources), then the integral in
\eqref{eqn7} becomes a function only of the signal detection threshold,
\(\snr^{*}\), and taking that to be a fixed parameter we can side-step the
divergence and say
\begin{equation}
\label{eqn9}
\text{\# sources}
   \propto \Dnl^{3}.
\end{equation}

\section{\(\tan^{-1}\ln\) Binning}
\label{appendix1}

In implementing the ranking statistic described in this article, one will
need to histogram randomly-drawn samples that span a large domain and whose
density varies greatly over that domain.  It is helpful to use non-uniform
binnings whose bin density is approximately inversely proportional to the
sample density so that the number of samples falling in each bin --- and
thus the error from counting fluctuations --- is approximately constant.

The functions described by the binned samples that are encountered here
tend to be well described by simple polynomials in the logarithm of the
function and the logarithm of the variable (see, for example, any of the
functions depicted in \figref{fig4}, \figref{fig3}, or \figref{fig5}), and
so binnings that are uniform in the logarithm of the variable would seem to
be convenient.  Unfortunately, in almost every case the variable is valid
from 0 to \(\infty\), and so an infinite number of uniform-in-the-logarithm
bins would be required to define the function everywhere.  To address this,
we make use of bins that are uniform in the arctangent of the logarithm of
the variable.  The arctangent function is approximately linear near 0 and
so by choosing an appropriate translation and scaling one obtains a binning
that is approximately logarithmic over some chosen range of values, but
that spans a domain from 0 to \(+\infty\) with a finite number of bins.

An \(\tan^{-1}\ln\) binning can be defined by three parameters:
\(x_{\mathrm{lo}}\) and \(x_{\mathrm{hi}}\), the low and high roll-offs
defining the range of approximately logarithmic binning, and \(n\), the
total number of bins.  The bin boundaries are given by
\begin{align}
x_{k}
   = \exp \left[
   \delta \frac{2}{\pi}
   \tan \left( \frac{\pi}{n} k - \frac{\pi}{2} \right)
   + \ln \bar{x}
   \right],
   && 0 \leq k \leq n,
\end{align}
where \(\ln \bar{x} = (\ln x_{\mathrm{hi}} + \ln x_{\mathrm{lo}}) / 2\),
and \(\delta = (\ln x_{\mathrm{hi}} - \ln x_{\mathrm{lo}}) / 2\).

One must be careful evaluating this function numerically, but, still,
except for binnings with very few bins, one nearly always finds that
underflows and overflows prevent the representation of the first few and
last few bin boundaries using double-precision floating-point numbers,
making the first few and last few bins appear to have identical upper and
lower boundaries.  We simply discard those bins, therefore generally the
number of useful bins is slightly less than \(n\).

For example, if \(x_{\mathrm{lo}} = 10\), \(x_{\mathrm{hi}} = 100\), and
\(n = 10\), the bin boundaries are 0, 3.314, 11.53, 18.57, 24.92, 31.62,
40.13, 53.86, 86.72, 301.8, \(\infty\).  The factors separating adjacent
boundaries are \(\infty\), 3.48,  1.61,  1.34,  1.27, 1.27, 1.34,  1.61,
3.48, \(\infty\).

\bibliography{references}
\end{document}